\documentclass[a4paper,10pt]{article}
\usepackage[scaled]{helvet}              
\usepackage[T1]{fontenc}                 
\usepackage{geometry}
\usepackage{setspace}                 
\usepackage{exscale}                  
\usepackage{a4wide}                   
\usepackage{framed}                   
\usepackage{subscript}                
\usepackage{authblk}                  
\usepackage{appendix}                 
\usepackage{colordvi}                 
\usepackage{enumitem}                 
\usepackage{graphicx}
\usepackage{float}
\usepackage{array}     
\usepackage[american]{babel}
\usepackage{fancyhdr}                   
\usepackage[colorlinks=true,linkcolor=black,citecolor=black,filecolor=black,urlcolor=black,hypertexnames=false]{hyperref}
\usepackage{aliascnt}      
\usepackage{cleveref}      
\usepackage[babel,german=guillemets]{csquotes}
\usepackage[backend=bibtex,style=numeric-comp,useprefix=true,hyperref=true,firstinits=true]{biblatex}
\bibliography{../../../Import/References/references.bib}
\usepackage[fleqn]{amsmath} 
\usepackage{mathtools}      
\usepackage{amssymb}        
\usepackage{latexsym}       
\usepackage{dsfont}         
\numberwithin{equation}{section}      
\input{../../../Import/Environments/thm-environments.tex}
\input{../../../Import/Macros/mathmacros-global.tex}
\newcommand*{\tx}{(t,x)}

\newcommand*{\constBoltz}{\ensuremath{ k_b }}           
\newcommand*{\constCharge}{\ensuremath{ e_0 }}          
\newcommand*{\constEL}{\ensuremath{ \epsilon_0 }}       
\newcommand*{\temp}{\ensuremath{ T }}                                       
\newcommand{\fieldF}{\ensuremath{ \vecu }}                                  
\newcommand*{\chargeEL}[1][]{\ensuremath{ \rho_{f{#1}} }}                   
\newcommand{\potEL}{\ensuremath{ \Phi }}                                    
\newcommand{\fieldEL}{\ensuremath{ \vecE }}                                 
\newcommand*{\ml}[1][l]{\ensuremath{ m_{{#1}} }}                                                    
\newcommand*{\Dl}[1][l]{\ensuremath{ d_{{#1}} }}                                                    
\newcommand*{\zl}[1][l]{\ensuremath{ z_{{#1}} }}                                                    
\newcommand{\pressHydrl}[1][l]{ p_{{#1}} }                                               
\newcommand{\rl}[1][l]{\ensuremath{ r_{{#1}} }}                                          
\newcommand{\Rl}[1][l]{\ensuremath{ R_{{#1}} }}                                          
\newcommand{\nl}[1][l]{\ensuremath{ n_{{#1}} }}                             
\newcommand{\yl}[1][l]{\ensuremath{ y_{{#1}} }}                              
\newcommand{\rol}[1][l]{\ensuremath{ \rho_{{#1}} }}                          
\newcommand{\slj}[1][l]{\ensuremath{ s_{{#1}j} }}                                        
\newcommand{\kfj}[1][j]{\ensuremath{ k^f_{{#1}} }}                                       
\newcommand{\kbj}[1][j]{\ensuremath{ k^b_{{#1}} }}                                       
\newcommand{\Rfjmal}[1][i]{\ensuremath{ \kfj \prod_{\slj[{#1}]<0} \yl[i]^{-\slj[{#1}]} }}   
\newcommand{\Rbjmal}[1][i]{\ensuremath{ \kbj \prod_{\slj[{#1}]>0} \yl[i]^{ \slj[{#1}]} }}   
\newcommand{\Rjmal}[1][i]{ \ensuremath{ \Rfjmal[i] - \Rbjmal[i] }}                       
\newcommand{\cpos}{\rol[]^+}
\newcommand{\ypos}{\yl[]^+}
\newcommand{\Dpos}{\Dl[]^+}
\newcommand{\zpos}{\zl[]^+}
\newcommand{\spos}{s^+}
\newcommand{\mpos}{\ml[]^+}
\newcommand{\cneg}{\rol[]^-}
\newcommand{\Dneg}{\Dl[]^-}
\newcommand{\zneg}{\zl[]^-}
\newcommand{\mneg}{\ml[]^-}
\newcommand{\capos}{\rol[a]^+}
\newcommand{\Dapos}{\Dl[a]^+}
\newcommand{\zapos}{\zl[a]^+}
\newcommand{\sapos}{s_a^+}
\newcommand{\mapos}{\ml[a]^+}
\newcommand{\csol}{\rol[s]}
\newcommand{\Rcoag}{\Rl[]^{coag}}
\input{../../../Import/Environments/environments.tex}
\let\sim~ 
\begin{document}
%
%
\title{Modeling and simulation of coagulation according to DLVO-theory in a continuum model for electrolyte solutions}
\author[1]{Matthias Herz}
\author[1]{Peter Knabner}
\renewcommand\Affilfont{\itshape\small}
\affil[1]{Department of Mathematics, University of Erlangen-N\"urnberg, Cauerstr. 11, D-91058 Erlangen, Germany \newline
          Fax: +49 +9131 85-67225; Tel: +49 +9131 85-67238; E-mail: matthias.herz@fau.de}
\date{\today}
\maketitle
%
%
\begin{abstract}
This paper presents a model of coagulation in electrolyte solutions. In this paper, the coagulation process is modeled according to DLVO-theory, which is an atomistic theory.
On the other hand, we describe the dynamics in the electrolyte solutions by the \pnp, which is a continuum model. The contribution of this paper is to include the atomistic
description of coagulation based on DLVO-theory in the continuum \pnp. Thereby, we involve information from different spatial scales.
For this reason, the presented model accounts for the short-range interactions and the long-range interactions, which drive the coagulation process.
Furthermore, many-body effects are naturally included as the resulting model is a continuum model.
\\[2.0mm]
\end{abstract}
\fancyhead[L]{Introduction}
\section{Introduction}
%
In multicomponent solutions, an ubiquitous process is the formation of particle clusters, which is commonly referred to as coagulation. Particle clusters arise when particles of the same chemical species
collide and afterwards stick together due to attractive particle interactions. The prototypical example of attractive particle interactions are ever present \vdW\ interactions. These attractive interactions originate from
electromagnetic fluctuations, 
cf. \cite{BuhmannWelsch, DaviesNinham, LifshitzLandau-book9, Ninham-book, Parsegian-book, Russel-book, van-Kampen}.
Moreover, \vdW\ interactions are short-range interactions, which are active over a few nanometers. Thus, \vdW\ interactions are solely relevant for particles close to contact.
On the other hand, for particles far from contact the driving force behind collisions is thermal motion and for charged particles additionally electric drift motion along the field lines
of a long-range electric field. However, in particular for electrically charged particles there is an additional short-range electric repulsion between equally charged particles,
which counteracts attractive \vdW\ interactions. Thus, in electrolyte solutions coagulation is caused by the interplay of long-range thermal motion, long-range electric fields,
short-range attractive \vdW\ interactions, and short-range electric repulsion of charged particles of the same chemical species. More precisely, \vdW\ interactions, moderate thermal motion,
and moderate electric drift motion trigger the coagulation process, whereas strong thermal motion, strong electric drift motion, and electrostatic repulsion prevent the particles from coagulation.
\par
Furthermore, coagulation is the central process, which determines the stability properties of electrolyte solutions. Here, an electrolyte solution is stable,
if dissolved particles remain dissolved, whereas an electrolyte solution is unstable, if dissolved particles form clusters, i.e., if the particles coagulate.
\medskip
\par
Classically, coagulation in electrolyte solutions is explained by the so-called DLVO-theory, which is at the heart of colloid science and electrochemistry,
cf. \cite{Elimelech-book, Hunter-book, Israelachvili-book, LyklemaEtAl, Lyklema-book4, Masliyah-book, NinhamDLVO, Probstein-book}. This short-range theory describes the
interplay of short-range attractive \vdW\ forces and short-range electrostatic repulsion in terms of a pair potential. Here, \enquote{pair potential} refers to an energy potential,
that describes for two particles the energy of interaction as function of distance. This means, that DLVO-theory explicitly resolves single particles and thus, is an atomistic theory.
However, DLVO-theory solely captures the energetic picture of coagulation, i.e., whether coagulation is an energetic favorable process or not. Hence, informations about coagulation kinetics
are absent in DLVO-theory. Regarding the dynamics of coagulation, classical models are the Smoluchowski coagulation equation or, more generally, population balance equations,
cf. \cite{AbelEtAl, Amann2000, BurgerEtAl-Aggration-2009, capasso_aggr, ChenElimelech, Dhont-book, HounslowEtAl, Hunter-book, IslamEtAl, Kahlweit, KobayashiEtAl, LattuadaEtAl-1, LattuadaEtAl-2, Masliyah-book, PengEtAl, RamkrishnaReview, SandersEtAl, Smorodin}.
Thus, a comprehensive model of coagulation has to include both, the energetic picture, and the kinetic picture.
\medskip
\par
In this paper, we describe the interplay of long-range thermal motion (=diffusion) and long-range electric drift motion (=electro diffusion) by means of the \nspnp.
This system is a thermodynamically consistent and classical continuum model for electrolyte solutions.
However, as the characteristic feature of continuum models is to simultaneously track a large number of particles by means of particle concentrations, these models do not resolve single particles
and thus, continuum models have no access to atomistic DLVO-theory. This is a serious restriction, as coagulation of particles occurs through the interplay of long-range interactions governed by the
\nspnp\ and short-range interactions governed by DLVO-theory. In this paper, we derive a model for coagulation in electrolyte solutions, which fully captures the process of coagulation in the sense
that we successfully combine the continuum \nspnp\ with the atomistic DLVO-theory. Thereby, we obtain a micro-macro model, which naturally accounts for the short-range picture, the long-range picture,
the energetic picture, and the kinetic picture.
\medskip
\par
The rest of this paper is organized as follows: Firstly, we present the governing equations in \cref{sec:coagES-start}. Next, in \cref{sec:coagES-kinetics}, we propose an ansatz for coagulation kinetics,
and we show how to include DLVO-theory in a continuum model. Finally, in \cref{sec:coagES-model}, we present the resulting model and some numerical simulations.
%
%
\section{The general governing Equations}\label{sec:coagES-start}
We adopt the \nspnp\ as the governing equations for the considered electrolyte solutions. Note, that this system is thermodynamically consistent (especially the thermodynamical consistency of this system, we investigate in a forthcoming paper)
model for isothermal, incompressible electrolyte solutions with an electrically neutral solvent. More precisely, this system consists of the following set of equations.
\\[3.0mm]
\begin{subequations}
\textbf{1. Poisson's equation: } We have for the electric field~$\fieldEL=-\grad\potEL$, and $\potEL$ solves
\begin{align}\label{eq:coagES-start-nspnpPoisson}
  -\grad\cdot(\eps_r\grad\potEL) = \frac{1}{\constEL}~ \chargeEL \qquad\text{with }\quad\chargeEL=\sum_l\frac{\constCharge\zl}{\ml}\rol~.
\end{align}
Here, $\constEL$ is the electric permittivity, $\eps_r$ the relative permittivity of the medium, $\chargeEL$ is the free charge density, $\constCharge$ the elementary charge.
For $l\in\{1,\ldots,L\}$, $\rol$ is the mass density of the $l$th solute, $\ml$ the molecular mass of the $l$th solute, and $\zl$ the valency of the $l$th solute. Furthermore, we have indexed the chemical species
such the the solvent is the $L$th chemical species. As we assume an electrically neutral solvent, we have $\zl[L]=0$.\\[2.0mm]
\textbf{2. Nernst--Planck equations: } For $l\in\cbrac{1,\ldots,L-1}$, we have
\begin{align}\label{eq:coagES-start-nspnpMassBalance}
   \dert\rol + \grad\cdot\brac{ \rol\fieldF -\Dl\grad\rol - \frac{\constCharge\Dl\zl}{\constBoltz\temp}\,\rol\fieldEL } ~=~\rl~,
\end{align}
Here, $\Dl$ is the diffusion coefficient of the $l$th solute, $\constBoltz$ the Boltzmann constant, and $\temp$ the constant temperature.
Furthermore, the ansatzes for the mass production rates~$\rl$ are given by mass action kinetics, cf. \cite{Atkins-book, EckGarckeKnabner-book, SmithMissen-book, PrigogineKondepudi-book, Upadhay-book}.
Note, that we deal with incompressible mixtures. This is modeled by a constant total density~$\rol[]:=\sum_l\rol \equiv const$, which leads to the following incompressibility
constraint~\eqref{eq:coagES-start-nspnpMassBalanceTot}. cf. \cite{DeGrootMazur-book, oden-book, Madja-book}. Thus, we obtain the solvent concentration~$\rol[L]$ by $\rol[L]= \rol[] - \sum_l^{L-1}\rol$,
cf.~\cite{DeGrootMazur-book, PrigogineKondepudi-book}.
\\[2.0mm]
\textbf{3. Navier--Stokes equations: } For the barycentric momentum density~$\rol[]\fieldF$ holds
\begin{align}
  & \grad\cdot\fieldF = 0~,                                                                                                                 \label{eq:coagES-start-nspnpMassBalanceTot}\\
  &\rol[] \dert\fieldF + \rol[]\grad\cdot\brac{\fieldF\otimes\fieldF} = -\grad\pressHydrl[] + 2\eta\Delta\fieldF + \chargeEL\fieldEL ~.   \label{eq:coagES-start-nspnpMomBalanceTot}
\end{align}
Here, $\fieldF$ is the barycentric velocity field and $\pressHydrl[]$ the pressure of the mixture.
\end{subequations}
\medskip
\par
To introduce the new coagulation model within the framework of \nspnp{s}, we subsequently reduce the
preceding equations~\eqref{eq:coagES-start-nspnpPoisson}--\eqref{eq:coagES-start-nspnpMomBalanceTot} to a setting, which allows to concentrate
on coagulation effects. More precisely, we henceforth confine ourselves to the following setting:
\begin{enumerate}[align=left, leftmargin=*, topsep=2.0mm, itemsep=0.0mm, label={(M\arabic*)}, start=1]
 \item We consider electrolyte solutions with four components, i.e, we have $L=4$.%
       \label{MODEL:assump1}
 \item We suppose, the four components are given by the neutral solvent~$\rol[s]:=\rol[4]$, a positively charged chemical species~$\cpos:=\rol[1]$, and a negatively charged chemical species~$\cneg:=\rol[2]$.
       Furthermore, we assume that solely the positively charged chemical species can coagulate and form aggregates. To resolve these aggregates, we introduce an additional chemical species~$\capos:=\rol[3]$.%
       \label{MODEL:assump2}
 \item Henceforth, we label the quantities for $\cpos$ by the superscript~${}^+$, the quantities for $\cneg$ by~${}^-$, the quantities for $\capos$ by~${}^+_a$,
       and the quantities for $\csol$ by the subscript~${}_s$.%
       \label{MODEL:assump3}
 \item We assume, the aggregates are subject to gravitational forces. Hence, we have with the gravitational acceleration field~$\vecg\sim[m/s^2]$
       the additional term~$\frac{\Dl[3]\ml[3]}{\constBoltz\temp}\,\capos\vecg\sim[kg/(m^2s)]$ in the mass flux for $\capos$.%
       \label{MODEL:assump4}
 \item We suppose symmetric charges~$\zpos=z=-\zneg$, and $\rl[-]=0$.%
       \label{MODEL:assump5}
 \item We assume, that the single reactive mechanism within the mixture is the coagulation of the positively charged chemical species, which can be described by a
       reaction rate~$\Rcoag$. Hence, the mass production rate~$\rl[]^+$ reads as $\rl[]^+=\mpos\spos\Rcoag$. Similarly the mass production rate~$\rl[a]^+$ is given by $\rl[a]^+=\mapos\sapos\Rcoag$,
       cf. \cite{BotheDreyer, DeGrootMazur-book, PrigogineKondepudi-book}%
       \label{MODEL:assump6}
\end{enumerate}
Thus, for the rest of this paper the governing equations are given by
\\[3.0mm]
\begin{subequations}
\textbf{1. Poisson's equation: } We have for the electric field~$\fieldEL=-\grad\potEL$, and $\potEL$ solves
\begin{align}\label{eq:coagES-start-poisson}
  -\grad\cdot(\eps_r\grad\potEL) = \frac{1}{\constEL}~ \chargeEL \qquad\text{with }\quad\chargeEL=\sum_l\frac{\constCharge\zl}{\ml}\rol=\frac{z\cpos}{\mpos}-\frac{z\cneg}{\mneg}+\frac{\zapos\capos}{\mapos}~.
\end{align}
\textbf{2. Nernst--Planck equations: } We have
\begin{flalign}
 &  \dert\cpos   + \grad\cdot\!\brac{ \cpos\fieldF  -\Dpos\grad\cpos   - \frac{\constCharge\Dpos z}{\constBoltz\temp}\,\cpos\fieldEL }   = \mpos\spos\Rcoag~,&   \label{eq:coagES-start-massBalance1} \\
 &  \dert\cneg   + \grad\cdot\!\brac{ \cneg\fieldF  -\Dneg\grad\cneg   + \frac{\constCharge\Dneg z}{\constBoltz\temp}\,\cneg\fieldEL }   = 0~,& \label{eq:coagES-start-massBalance2}\\
 &  \dert\capos  + \grad\cdot\!\brac{ \capos\fieldF -\Dapos\grad\capos - \frac{\constCharge\Dapos\zapos}{\constBoltz\temp}\,\capos\fieldEL +\frac{\Dapos\ml[a]^+}{\constBoltz\temp}\,\capos\vecg} = \mapos\sapos\Rcoag\,. & \label{eq:coagES-start-massBalance3}
\end{flalign}
Here, the solvent concentration~$\csol$ is obtained by $\csol= \rol[] - \cpos -\cneg -\capos$.
\\[3.0mm]
\textbf{3. Navier--Stokes equations: } For the barycentric momentum density holds
\begin{align}
  & \grad\cdot\fieldF = 0~,                                                                                                                 \label{eq:coagES-start-massBalanceTot}\\
  &\rol[] \dert\fieldF + \rol[]\grad\cdot\brac{\fieldF\otimes\fieldF} = -\grad\pressHydrl[] + 2\eta\Delta\fieldF + \chargeEL\fieldEL ~.   \label{eq:coagES-start-momBalanceTot}
\end{align}
\end{subequations}
\medskip
In the next section, we derive a new ansatz for the rate function~$\Rl[]^{coag}$, which models the process of coagulation.
%
%
\section{Coagulation Kinetics including DLVO-theory}\label{sec:coagES-kinetics}
%
First of all, we note that the reaction rates in equations~\eqref{eq:coagES-start-massBalance1}--\eqref{eq:coagES-start-massBalance3} are reaction rates according to mass action law,
cf. \cite{Atkins-book, EckGarckeKnabner-book, SmithMissen-book, PrigogineKondepudi-book, Upadhay-book}. More precisely, from these references we recall that provided the stoichiometry
of a chemical reaction reads as
\begin{align*}
 \tilde{s}_1[C_1] + \tilde{s}_2[C_2] ~\stackrel{}{\begin{subarray}{l} -\!\!\!-\!\!\!-\!\!\!-\!\!\!-\!\!\!\!\!\rightharpoonup \\[-1.5mm] \leftharpoondown\!\!\!-\!\!\!-\!\!\!-\!\!\!- \end{subarray}}~ \tilde{s}_3[C_3]~,
\end{align*}
the corresponding reaction rate function~$\Rl[]$ is given with the mass fractions~$\yl$ and the stoichiometric coefficients~$s_1:=-\tilde{s}_1$, $s_2:=-\tilde{s}_2$, $s_3:=\tilde{s}_3$ by
\begin{align}\label{eq:coagES-kinetics-mal}
 \Rl[] = \Rl[]^f - \Rl[]^b = \kfj[] \yl[1]^{-s_1} \yl[2]^{-s_2} - \kbj[] \yl[3]^{s_3}~.
\end{align}
According to this equation the forward reaction takes place, if both reactants are present. Hence, the above stoichiometry implicitly contains the assumption that the single barrier,
which hinders the reactions is the presence of the reactants. However, in some situations other barriers such as certain energy barriers truly exist.
Provided we formulate such a barrier with a function~$g$ as an inequality constraint~$g>0$, the stoichiometry now is given by
\begin{align}\label{eq:coagES-kinetics-stoichiometry}
 \tilde{s}_1[C_1] + \tilde{s}_2[C_2]
 ~~\stackrel{g>0~}{\begin{subarray}{l} -\!\!\!-\!\!\!-\!\!\!-\!\!\!-\!\!\!-\!\!\!-\!\!\!-\!\!\!\!\!\rightharpoonup \\[-1.5mm] \leftharpoondown\!\!\!-\!\!\!-\!\!\!-\!\!\!-\!\!\!-\!\!\!-\!\!\!- \end{subarray}}~ \tilde{s}_3[C_3]~.
\end{align}
Here, the condition~$g>0$ above the arrows indicates that the reaction solely takes place if we have~$g>0$. Thus, besides the presence of the reactants the reaction rate is triggered
by a barrier function~$H(g)$, which is given with the Heaviside function~$H$ by
\begin{align*}
 \tx \mapsto H(g\tx) = \begin{cases} 1, & \text{ if } g\tx>0~, \\ 0, & \text{ if } g\tx\leq 0~. \end{cases}
\end{align*}
Thus, the rate functions~$\Rl[]$ for reactions, which include inequality constraints, are given by
\begin{align}\label{eq:coagES-kinetics-rateFuncExtended}
 \Rl[] = \underbrace{H(g)}_{\begin{subarray}{l} \text{barrier} \\ \text{function} \end{subarray}} \,
         \underbrace{\Rl[]^{kin}}_{\begin{subarray}{l} \text{reaction} \\ \text{kinetics} \end{subarray}} \
       = \begin{cases} \Rl[]^{kin}, & \text{ if } g>0~, \\ 0, & \text{ if } g\leq 0~. \end{cases}
\end{align}
Here, the barrier function~$H(g)$ triggers the reaction according to the given inequality constraint and $\Rl[]^{kin}$ describes the reaction kinetics. Thus, the mass action law rate function~$\Rl[]$
corresponding to \eqref{eq:coagES-kinetics-stoichiometry} is given with \eqref{eq:coagES-kinetics-mal} and \eqref{eq:coagES-kinetics-rateFuncExtended} by
\begin{align*}
 \Rl[]  = H(g) \, \brac{\kfj[] \yl[1]^{-s_1} \yl[2]^{-s_2} - \kbj[] \yl[3]^{s_3}}
        = \begin{cases} \kfj[] \yl[1]^{-s_1} \yl[2]^{-s_2} - \kbj[] \yl[3]^{s_3}, & \text{ if } g>0~, \\ 0, & \text{ if } g\leq 0~. \end{cases}
\end{align*}
Note, that reaction rate functions including the Heaviside function have previously been introduced in a different context in \cite{KnabnerDuijnHengst}. Subsequently, we introduce the coagulation rate~$\Rcoag$
exactly as reaction rate of the above type. More precisely, to include in the coagulation rate~$\Rcoag$ the energetic picture according to DLVO-theroy as well as the kinetic picture of coagulation,
we assume for $\Rcoag$ the ansatz
\begin{align}\label{eq:coagES-kinetics-ansatzCoagRateFunction}
 \Rcoag = H\!\brac{g^{DLVO}}~ \Rl[]^{coag,kin}~.
\end{align}
Here, $g^{DLVO}$ has to capture the energy barrier according to DLVO-theory and $\Rl[]^{coag,kin}$ the coagulation kinetics. In the rest of this section,
we firstly derive a formula for $\Rl[]^{coag,kin}$, and secondly a formula for $g^{DLVO}$.
\medskip
\par
In this passage, we introduce the coagulation kinetics. For that purpose, we assume that the aggregates~$[C^+_a]$ are made of doublets of dissolved particles~$[C^+]$.
This leads us to the stoichiometry
\begin{align}\label{eq:coagES-kinetics-ansatzCoagKinetics}
 [C^+] + [C^+] ~-\!\!\!-\!\!\!\!\!\rightharpoonup~ [C^+_a]
 \qquad\Equivalent\qquad
 2[C^+] ~-\!\!\!-\!\!\!\!\!\rightharpoonup~ [C^+_a]~.
\end{align}
As this stoichiometry is a crucial part of the proposed model, we now clarify this ansatz.
\begin{enumerate}[align=left, leftmargin=*, topsep=2.0mm, itemsep=0.0mm, label={(\roman*)}, start=1]
 \item The arrow \enquote{$\rightharpoonup$} indicates that we consider coagulation as an irreversible reaction. This comes from the fact that we suppose the
       aggregates to be solid aggregates. For solids the dissolution rate just depends on the intensity of the thermal motion of the particles within
       the solid structure, i.e., the dissolution rate is a function of temperature, cf. \cite{Atkins-book, EckGarckeKnabner-book}. As we consider isothermal
       electrolyte solutions, the dissolution rate is constant. Henceforth, we assume this constant to be zero.
 \item The assumption that the aggregates are made of doublets leads to a quadratic reaction rate. This corresponds to the Smoluchowski coagulation equation,
       which is based on quadratic coagulation kernels, cf. \cite{Dhont-book, Hunter-book, Masliyah-book}.
 \item Note, that the aggregates~$[C^+_a]$ are of the same chemical species as~$[C^+]$. Furthermore, in \ref{MODEL:assump4} we supposed that the aggregates are subject
       to gravity. In the framework of population balance equations this means, we have separated the positively charged particles into two size classes.
       The first size class contains the particles~$[C^+]$, which are sufficiently small such that gravitational effects can be neglected, and such that we can consider
       this particles to be dissolved. The second size class contains the aggregates~$[C^+_a]$, which are sufficiently large such that they are subject to gravity, and
       such that we can consider the particles of this size class to be solid aggregates.
 \item The preceding interpretation of the model in terms of size classes reveals that only a fraction of doublets~$2[C^+]$ leads to an aggregate~$[C^+_a]$ in the second size class.
       More precisely, the particles~$[C^+]$ can from doublets, doublets of doublets, etc. Only after repeatedly forming doublets of doublets, the resulting aggregates have grown
       large enough to belong to the second size class~$[C^+_a]$. Hence, the stoichiometry~\eqref{eq:coagES-kinetics-ansatzCoagKinetics} includes a certain latency due to repeatedly formation of doubles.
       Thus, the forward reaction rate constant~$\kfj[]$ must account for this latency factor.
 \item From \eqref{eq:coagES-kinetics-ansatzCoagKinetics}, we know that the stoichiometric coefficients are given by $\spos=-2$, $\sapos=1$, cf. \cite{Atkins-book, EckGarckeKnabner-book, SmithMissen-book, Upadhay-book}.
       Thus, we define the valencies by $\zapos=2\zpos$, and the molecular masses by $\mapos=2\mpos$.
       This choices ensure the conservation of mass, i.e., $\mpos\spos\Rl[]^{coag} +\mapos\sapos\Rl[]^{coag} = 0$,
       and the conservation of charges, i.e., $\constCharge\zpos\spos\Rl[]^{coag} +\constCharge\zapos\sapos\Rl[]^{coag} = 0$, cf. \cite{BotheDreyer, DeGrootMazur-book, PrigogineKondepudi-book}.
\end{enumerate}
Altogether, we deduce from \eqref{eq:coagES-kinetics-ansatzCoagKinetics} that the coagulation kinetics are given by
\begin{align}\label{eq:coagES-kinetics-coagKinetics}
 R^{coag,kin} = \kfj[](\ypos)^2.
\end{align}
\par
As to the energy barrier according to DLVO-theory, we first recall that DLVO-theory describes the interplay between short-range electrostatic repulsion
and short-range attractive \vdW\ interactions, cf. \cite{Hunter-book, Masliyah-book, Lyklema-book4}. Van der Waals interactions originate from electromagnetic fluctuations,
cf. \cite{BuhmannWelsch, DaviesNinham, LifshitzLandau-book9, Ninham-book, Parsegian-book, Russel-book, van-Kampen}, and thus, the short range of \vdW\ interactions
inherently lies in the nature of \vdW\ interactions. Opposed to this, electrostatic interactions are commonly considered to be long-range interactions. However, in electrolyte
solutions the charge of a single particle is screened over a very short distance by oppositely charged particles. Consequently, owing to this screening, the charge of a particle
is active only in a very small neighborhood, which is commonly referred to as the electric double layer (EDL), cf. \cite{Hunter-book, Israelachvili-book, Masliyah-book, Probstein-book, Russel-book}.
Hence, the electrostatic repulsion between equally charged particles takes place as short-range interaction between their EDLs. More precisely, two equally charged particles
only electrically interact provided their EDLs overlap. As a typical thickness of a EDL is 10 nm, cf. \cite{Hunter-book, Masliyah-book}, particles have to approach very close
for overlapping EDLs. This screening of electrostatic interactions to short-range EDL interactions in electrolyte solutions is the deeper reason why \vdW\ interactions become
active in the first place.
\par
Following the well-known derivations in \cite{Hunter-book, Israelachvili-book, Lyklema-book4, Masliyah-book, Probstein-book, Russel-book}, DLVO-theory can be condensed by means
of the so-called DLVO-potential~$V^{tot}$, which is given by
\begin{align*}
 V^{tot} = V^{el} + V^{vdw}~.
\end{align*}
Here, $V^{el}$ is the potential accounting for electrostatic repulsion, and $V^{vdw}$ the potential accounting for \vdW\ interactions. More precisely, $V^{el}$ describes the
electrostatic repulsion between two infinite plates separated at distance~$d$ and $V^{vdw}$ describes nonretarded \vdW\ interactions between two infinite plates separated
at distance~$d$. These potentials are given according to \cite[Chapter~12.6]{Hunter-book} by
\begin{align*}
  V^{el}(d)  = \frac{64 \nl[b]\constBoltz\temp}{\kappa}\, \exp(-\kappa d)
  \qquad\text{and}\qquad
  V^{vdw}(d) = -\frac{A}{12\pi d^2}~.
\end{align*}
Here, $\kappa^{-1}\sim[m]$ is the Debye length, $\nl[b]$ the number bulk concentration of the negatively charged counterions, and $A\sim[J]$ the Hamaker constant.
These formulas show, that the DLVO-potential~$V^{tot}=V^{tot}(d)$ is a pair potential, which expresses the interaction energy between two particles as function of
interparticle distance~$d$.
\begin{figure}[H]
 \begin{center}
   \includegraphics*[bb=0 40 650 700, scale=0.3]{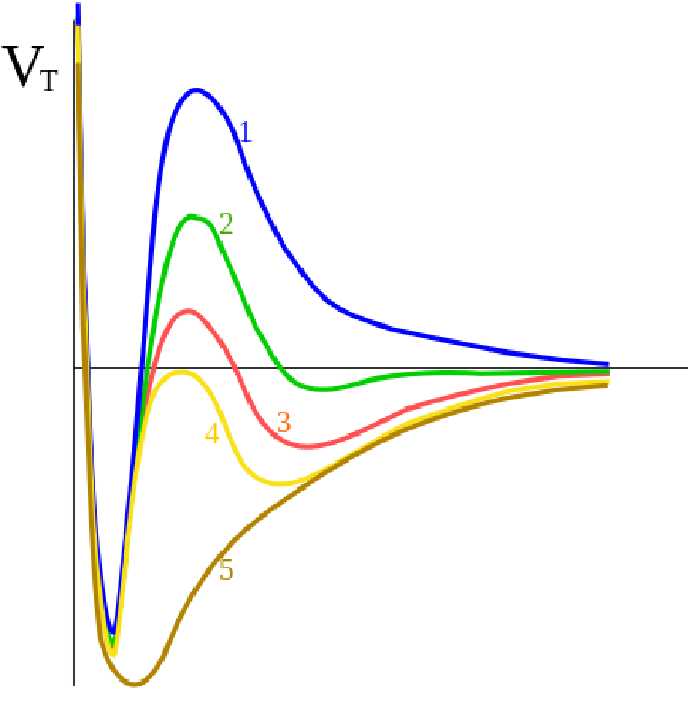} 
 \end{center}
 \vspace{-7.0mm}
 \caption{Ideal-typical plots of the DLVO-potential, cf. \cite{DLVO-figure}}\label{fig:coagES-dlvo}
\end{figure}
Furthermore, we have $V^{tot}<0$ in case of attractive interaction energies and $V^{tot}>0$ in case of repulsive interaction energies.
\cref{fig:coagES-dlvo} depicts some ideal-typical plots of the DLVO-potential. This figure reveals two characteristic features of the DLVO-potential.
Firstly, the DLVO-potential possesses an absolute minimum, which is the so-called primary minimum, cf. \cite{Hunter-book, Israelachvili-book, Masliyah-book, Probstein-book, Russel-book}.
The DLVO-potential~$V^{tot}$ reaches this primary minimum precisely in case of coagulation. Hence, from an energy minimizing perspective, coagulation is the energetically favorable state.
However, the second characteristic feature of the DLVO-potential~$V^{tot}$ is the energy barrier, which is depicted by the local maximum in the respective curves in \cref{fig:coagES-dlvo}.
This local maximum of the DLVO-potential is exactly the condensed description of the energy barrier due to electrostatic repulsion, which prevents the particles from coagulation.
In curve~5 this energy barrier is absent, as this curve solely depicts the attractive part~$V^{vdw}$. However, the essential observation from \cref{fig:coagES-dlvo} is that the local
maximum only leads to an energy barrier, if the value of the DLVO-potential~$V^{tot}$ at this point is positive, as positive values mean repulsion. Consequently, provided the DLVO-potetial~$V^{tot}$
is given by curve~4, we have no energy barrier. Moreover, as $V^{tot}$ has only nonpositive values along curve~4, this curve depicts a purely nonrepulsive potential~$V^{tot}$.
Thus, in this situation coagulation can unconditionally take place. Consequently, the succinct criterion for such a nonrepulsive situation is a double zero of $V^{tot}$, i.e.,
\begin{align*}
 V^{tot}=0=(V^{tot})^\prime~.
\end{align*}
Following the calculations in~\cite{Hunter-book}, we firstly solve for $(V^{tot})^\prime=0$. Thereby, we obtain with the Debye-length~$\kappa^{-1}$ the critical point~$x_*= 2/\kappa$.
Note, that in \cite{Hunter-book} the Debye-length~$\kappa^{-1}$ is defined by
\begin{align*}
 \kappa:=\brac{\frac{\constCharge^2\zl[]^2\nl[b]}{\eps_r\constEL\constBoltz\temp}}^{\frac{1}{2}}~.
\end{align*}
Substituting $x_*=2/\kappa$ and the preceding definition of $\kappa$ into $V^{tot}(x_*)=0$, yields together with $\zl[]^2=(\zneg)^2$ an equation, which can be solved for the number bulk concentration~$\nl[b]$.
Running through the calculations in \cite{Hunter-book} shows that this leads finally for the mass bulk concentration~$\rol[b]$ of the negatively charged particles to%
\footnote{Note that in \cite{Hunter-book} the formula for $c.c.c.$ is slightly different, as $c.c.c.$ is given in units $[mol/L]$, whereas we convert $c.c.c.$ to $[kg/m^3]$}
\begin{align*}
  \rol[b] := \mneg\nl[b] = 1000\mneg\frac{(4\pi\constEL)^3 0.107 \eps_r^3 (\constBoltz\temp)^5}{A^2(\constCharge\zneg)^6} =: c.c.c.
\end{align*}
The value of the right hand side is a given constant, which depends on several parameters. This constant is commonly referred to as the
critical coagulation concentration ($c.c.c.$). Hence, the succinct criterion for an energy barrier is
\begin{flalign}\label{eq:coagES-kinetics-dlvoCriterion}
 &~~\begin{array}{l c l c l}
     \rol[b] <    c.c.c. & \Hence & V^{tot}>0    \text{ at the local maximum } & \Hence &  \text{ energy barrier,}    \\
     \rol[b] \geq c.c.c. & \Hence & V^{tot}\leq0 \text{ at the local maximum } & \Hence &  \text{ no energy barrier.}
    \end{array} &
\end{flalign}
Here, it is essential that the inequality $\rol[b]\gtreqqless c.c.c.$ involves a macroscopic quantity~$\rol[b]$
on the left-hand side and a microscopic quantity~$c.c.c.$ on the right-hand side. More precisely, we already mentioned that DLVO-theory
captures the interplay of \vdW\ attraction and electrostatic repulsion of two particles with overlapping EDLs. Thus, DLVO-theory draws a
detailed microscopic picture on spatial scales, which are not resolved in macroscopic continuum models. As the~$c.c.c.$ is obtained by
collecting all parameters of the DLVO-potential, the microscopic picture is condensed in this constant. On the other hand,
within DLVO-theory bulk concentrations describe the concentration values outside the EDL regime. Thus, exactly these
bulk concentrations can be interpreted as the macroscopic densities, which enter continuum mechanical models. Furthermore, from a
microscopic point of view, bulk concentrations are given constants, whereas from a macroscopic point of view, bulk concentrations are
variable densities. Thus, the connection between microscopic DLVO-theory and macroscopic continuum models is to identify the bulk
concentration~$\rol[b]$ of the negatively charged particles with the macroscopic concentration~$\cneg$.
Thereby, we obtain from \eqref{eq:coagES-kinetics-dlvoCriterion} the criterion for coagulation
\begin{align*}
 \cneg \geq c.c.c. ~~\Hence~~ \begin{array}{l} \text{no microscopic energy barrier} \\ \text{according to DLVO-theory} \end{array}~~\Hence~~\text{coagulation.}
\end{align*}
Thus, we now define the barrier function~$g^{DLVO}$ from \eqref{eq:coagES-kinetics-ansatzCoagRateFunction} by
\begin{align}\label{eq:coagES-kinetics-dlvoBarrierFunc}
 g^{DLVO} = \cneg - c.c.c.
\end{align}
Consequently, according to \eqref{eq:coagES-kinetics-stoichiometry} and \eqref{eq:coagES-kinetics-ansatzCoagKinetics} we propose for coagulation the following stoichiometry
\begin{align*}
 [C_+] + [C_+] ~\stackrel{\cneg-c.c.c.>0}{-\!\!\!-\!\!\!-\!\!\!-\!\!\!-\!\!\!-\!\!\!-\!\!\!-\!\!\!-\!\!\!-\!\!\!\!\!\rightharpoonup}~ [C_{+,a}]~.
\end{align*}
In conclusion, this leads us with \eqref{eq:coagES-kinetics-ansatzCoagRateFunction} , \eqref{eq:coagES-kinetics-coagKinetics},
and \eqref{eq:coagES-kinetics-dlvoBarrierFunc} to the coagulation rate function
\begin{align*}
 \Rl[]^{coag} = \kfj[] H(\cneg - c.c.c.) \brac{\ypos}^2~.
\end{align*}
%
%
\section{Mathematical Model for Coagulation}\label{sec:coagES-model}
First of all, we recall that the mixture density~$\rol[]$ is a given constant due to assumption~\ref{MODEL:assump6}. Thus, we transform $\Rl[]^{coag}$ with $\cpos=\rol[]\ypos$ to
\begin{align*}
 \Rl[]^{coag} = \underbrace{\rol[]^{-2}\kfj[]}_{=:\tilde{k}^f} H(\cneg - c.c.c.) \brac{\cpos}^2 =\tilde{k}^f H(\cneg - c.c.c.) \brac{\cpos}^2.
\end{align*}
For ease of readability, we henceforth again write ~$\kfj[]$ instead of $\tilde{k}^f$. By substituting this equation and the additional preceding observations into the model equations
from \cref{sec:coagES-start}, we finally obtain the proposed continuum model for coagulation in electrolyte solutions:
\\[3.0mm]
\begin{subequations}
\textbf{1. Poisson's equation: } We have for the electric field~$\fieldEL=-\grad\potEL$, and $\potEL$ solves
\begin{align}\label{eq:coagES-model-poisson}
  -\grad\cdot(\eps_r\grad\potEL) = \frac{1}{\constEL}~ \chargeEL \qquad\text{with }\quad \chargeEL=\sum_l\frac{\constCharge\zl}{\ml}\rol=\frac{z\cpos}{\mpos}-\frac{z\cneg}{\mneg}+\frac{z\capos}{\mpos}~.
\end{align}
\textbf{2. Nernst--Planck equations: } We have
\begin{flalign}
 &  \dert\cpos   + \grad\cdot\!\brac{ \cpos\fieldF  -\Dpos\grad\cpos   - \frac{\constCharge\Dpos z}{\constBoltz\temp}\,\cpos\fieldEL }   = -2\mpos \Rl[]^{coag}~,&   \label{eq:coagES-model-massBalance1} \\
 &  \dert\cneg   + \grad\cdot\!\brac{ \cneg\fieldF  -\Dneg\grad\cneg   + \frac{\constCharge\Dneg z}{\constBoltz\temp}\,\cneg\fieldEL }   = 0~,& \label{eq:coagES-model-massBalance2}\\
 &  \dert\capos  + \grad\cdot\!\brac{ \capos\fieldF -\Dapos\grad\capos - \frac{2\constCharge\Dapos z}{\constBoltz\temp}\,\capos\fieldEL +\frac{2\Dapos\mpos}{\constBoltz\temp}\,\capos\vecg} = 2\mpos \Rl[]^{coag}\,. & \label{eq:coagES-model-massBalance3}
\end{flalign}
Here, the coagulation rate function~$\Rl[]^{coag}$ is given by
\begin{align}\label{eq:coagES-model-reactionRate}
 \Rl[]^{coag} = \kfj[] H(\cneg - c.c.c.) \brac{\cpos}^2,
\end{align}
and the solvent concentration~$\csol$ is obtained by $\csol= \rol[] - \cpos-\cneg-\capos$.
\\[3.0mm]
\textbf{3. Navier--Stokes equations: } For the barycentric momentum density holds
\begin{align}
  & \grad\cdot\fieldF = 0~,                                                                                                                 \label{eq:coagES-model-massBalanceTot}\\
  &\rol[] \dert\fieldF + \rol[]\grad\cdot\brac{\fieldF\otimes\fieldF} = -\grad\pressHydrl[] + 2\eta\Delta\fieldF + \chargeEL\fieldEL ~.   \label{eq:coagES-model-momBalanceTot}
\end{align}
\end{subequations}
The mass production rates are subject to conservation of total mass and charges, cf. \cite{DeGrootMazur-book, EckGarckeKnabner-book, PrigogineKondepudi-book}.
Furthermore, this model is a closed, as we have $5+n$~equations~\eqref{eq:coagES-model-poisson}--\eqref{eq:coagES-model-massBalance3}, \eqref{eq:coagES-model-massBalanceTot}, \eqref{eq:coagES-model-momBalanceTot}
for the $5+n$ unknowns
\begin{align*}
 \brac{\potEL,\cpos,\cneg,\capos,\pressHydrl[],\fieldF}\in\setR^{5+n}~.
\end{align*}
Next, we note that in situations with vanishing barycentric flow, we apparently have $\fieldF \equiv 0$. Substituting this into equations \eqref{eq:coagES-model-massBalanceTot} and
\eqref{eq:coagES-model-momBalanceTot} shows that the Navier--Stokes equations reduce to $\grad\pressHydrl[] = \chargeEL\fieldEL$. Thus, in case of vanishing barycentric flow
the pressure~$\grad\pressHydrl[]$ is determined up to a constant by $\potEL$ and the concentrations~$\cpos$, $\cneg$, and $\capos$, as we have~$\fieldEL=-\grad\potEL$ and $\chargeEL$
defined in \eqref{eq:coagES-model-poisson}. Consequently, in these situations the vector of primal unknowns is given by
\begin{align*}
 \brac{\potEL,\cpos,\cneg,\capos}\in\setR^{4}~,
\end{align*}
and the model equations~\eqref{eq:coagES-model-poisson}--\eqref{eq:coagES-model-momBalanceTot} reduce to
\\[3.0mm]
\textbf{\pnp: }
\begin{subequations}
\begin{flalign}
 & -\grad\cdot(\eps_r\grad\potEL) = \frac{1}{\constEL}~ \chargeEL \qquad\text{with }\quad\chargeEL=\sum_l\frac{\constCharge\zl}{\ml}\rol=\frac{z\cpos}{\mpos}-\frac{z\cneg}{\mneg}+\frac{z\capos}{\mpos}~,& \label{eq:coagES-model-pnp1} \\
 &  \dert\cpos   + \grad\cdot\!\brac{ \cpos\fieldF  -\Dpos\grad\cpos   - \frac{\constCharge\Dpos z}{\constBoltz\temp}\,\cpos\fieldEL }   = -2\mpos \Rl[]^{coag}~,&   \label{eq:coagES-model-pnp2} \\
 &  \dert\cneg   + \grad\cdot\!\brac{ \cneg\fieldF  -\Dneg\grad\cneg   + \frac{\constCharge\Dneg z}{\constBoltz\temp}\,\cneg\fieldEL }   = 0~,& \label{eq:coagES-model-pnp3}\\
 &  \dert\capos  + \grad\cdot\!\brac{ \capos\fieldF -\Dapos\grad\capos - \frac{2\constCharge\Dapos z}{\constBoltz\temp}\,\capos\fieldEL +\frac{2\Dapos\mpos}{\constBoltz\temp}\,\capos\vecg} = 2\mpos \Rl[]^{coag}\,. & \label{eq:coagES-model-pnp4}
\end{flalign}
Here, the coagulation rate function~$\Rl[]^{coag}$ is given by
\begin{align}\label{eq:coagES-model-pnp5}
 \Rl[]^{coag} = \kfj[] H(\cneg - c.c.c.) \brac{\cpos}^2~.
\end{align}
\end{subequations}
This subsystem of the model equations~\eqref{eq:coagES-model-poisson}--\eqref{eq:coagES-model-momBalanceTot} still captures the interplay between coagulation, diffusion, and electric drift motion.
More precisely, coagulation, diffusion, and electric drift motion are precisely the essential processes in almost all systems, in which coagulation occur. Thus, the restriction to $\fieldF\equiv0$
and to equations~\eqref{eq:coagES-model-pnp1}--\eqref{eq:coagES-model-pnp5} still encompasses most of the relevant systems.
\medskip
\par
Finally, we numerically solve the model equations~\eqref{eq:coagES-model-pnp1}--\eqref{eq:coagES-model-pnp5} in two space dimensions. Thereby, we demonstrate that this model is capable to produce the
expected coagulation kinetics. Concerning the computational setup, we choose as time interval $I:=[0,60]$ and as spatial computational domain~$\Omega\in\setR^2$, the rectangle~$\Omega:=[0,1]\times[0,2]$.
The boundary~$\Gamma$ of this domain, we split into the upper boundary~$\Gamma^a$, the lower boundary~$\Gamma^b$, the left boundary~$\Gamma^l$, and the right boundary~$\Gamma^r$,
which are respectively defined by
\begin{align*}
 \Gamma^a:=[0,1]\times\{2\}, \quad \Gamma^b:=[0,1]\times\{0\}, \quad \Gamma^l:=\{0\}\times[0,2], \quad \Gamma^r:=\{1\}\times[0,2]~.
\end{align*}
To obtain a computable model, we equip the preceding system of equations with the following initial conditions
\begin{align*}
 \cpos(0,x)\equiv1 , \qquad \cneg(0,x)\equiv0, \qquad \capos(0,x)\equiv0~,
\end{align*}
and boundary conditions
\begin{align*}
  &\potEL = 0 \text{ on } \brac{\Gamma^a \cup \Gamma^b}\times I \qquad\text{and}\qquad \potEL = c^{neg}<0 \text{ on } \brac{\Gamma^u \cup \Gamma^l}\times I~, \\
  &\brac{ -\Dpos\grad\cpos   - \frac{\constCharge\Dpos z}{\constBoltz\temp}\,\cpos\fieldEL }\cdot\vecnu =0~ \text{ on } \Gamma\times I, \\
  &\brac{ -\Dneg\grad\cneg   + \frac{\constCharge\Dneg z}{\constBoltz\temp}\,\cneg\fieldEL }\cdot\vecnu = h^{in} \text{ on } \Gamma\times I,\\
  &\brac{ -\Dapos\grad\capos - \frac{2\constCharge\Dapos z}{\constBoltz\temp}\,\capos\fieldEL +\frac{\Dapos\ml[a]^+}{\constBoltz\temp}\,\capos\vecg} \cdot\vecnu=0 \text{ on } \Gamma\times I~.
\end{align*}
In particular, the inflow function~$h^{in}$ is given by
\begin{align*}
 h^{in}\tx = \begin{cases}  c^{in}\,\chi_{[5,10]}(t), & \text{ on } \Gamma^a\times I~, \\ 0, & \text{ on } \brac{\Gamma\backslash\Gamma^a}\times I~. \end{cases}
\end{align*}
Here, $c^{in}$ is a given constant and $\chi_{A}(t)$ is the characteristic function for any set~$A\in\setR$, i.e.,
\begin{align*}
  \chi_{A}(t) = \begin{cases}  1, & \text{ if } t\in A~, \\ 0, & \text{ else }~. \end{cases}
\end{align*}
For the following simulation, we choose purely academic values for the coefficients. More precisely, we set $c^{neg}=0.5$, $c^{in}=0.2$, $\Dpos=\Dneg=0.1$, $\Dapos=0.01$, $\zl[]=1$, $\mpos=0.5$,
$\constCharge/(\constBoltz\temp)=10$, $\eps_r=0.1$, $\constCharge/\constEL=0.8$, $\kfj[]=2.5$, $c.c.c=0.75$. Furthermore, as discretization parameters we use the time step size~$\tau=0.1$ and
the mesh fineness~$h=0.05$, cf. \cite{Quarteroni-book, Larson-FE-book, Knabner-FE-book}. This computational setup produces the following dynamics:
\begin{enumerate}[align=left, leftmargin=*, topsep=2.0mm, itemsep=0.0mm, label={(\roman*)}, start=1]
 \item We can imagine $\Omega$ as a cross section of an idealized container, which is closed at the bottom, on the left, and on the right. Furthermore, this container possesses a constant surface potential~$c^{neg}$
       on the left wall and right wall. This forces the positively charged particles to accumulate at $\Gamma^l$ resp. $\Gamma^r$, see \cref{fig:coagES-sim0008}. In this figure, the solutions of
       equations~\eqref{eq:coagES-model-pnp1}--\eqref{eq:coagES-model-pnp5} are plotted as follows: From left to right, we have the concentration profiles of $\cpos$, $\cneg$ $\capos$, and the electric potential~$\potEL$
       together with the electric field lines.
       \begin{figure}[H]
	 \begin{center}
	  \includegraphics*[bb=50 40 1200 420, scale=0.32]{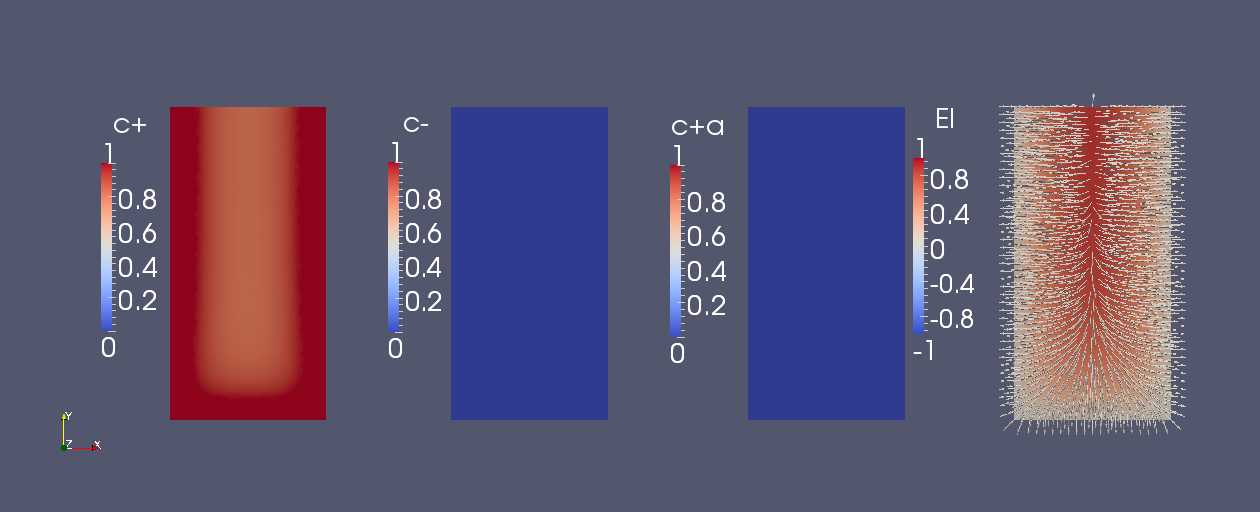} 
	 \end{center}
	 \vspace{-7.0mm}
	 \caption{Plot of the solutions at time $t=0.8$}\label{fig:coagES-sim0008}
       \end{figure}
 \item Initially, the only solute present is $\cpos$, and in the time interval~$[5,10]$, we have an inflow of $\cneg$ from above. According to \eqref{eq:coagES-model-pnp5}, this causes $\cpos$
       to coagulate, if we have $\cneg(t_0,x_0)>c.c.c.$, cf. \cref{fig:coagES-sim0089}. Here, the ordering of the solutions is identical to \cref{fig:coagES-sim0008}, and $c.c.c.$ is depicted by the white contour
       line in the second picture.
       \begin{figure}[H]
        \begin{center}
         \includegraphics*[bb=50 40 1200 420, scale=0.32]{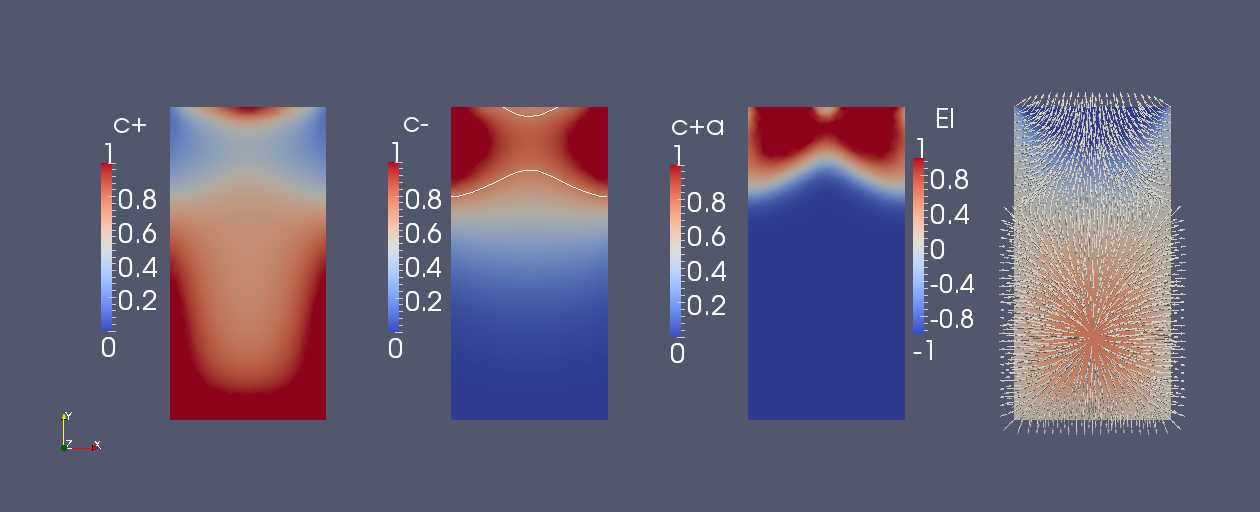} 
        \end{center}
        \vspace{-7.0mm}
        \caption{Plot of the solutions at time $t=8.9$}\label{fig:coagES-sim0089}
       \end{figure}
 \item The inflow of $\cneg$ stops at $t=10$. However, due to gravitational forces the aggregates continue to settle, cf. \cref{fig:coagES-sim0256}. Again, the solutions are ordered as in \cref{fig:coagES-sim0008}.
       \begin{figure}[H]
        \begin{center}
         \includegraphics*[bb=50 40 1200 420, scale=0.32]{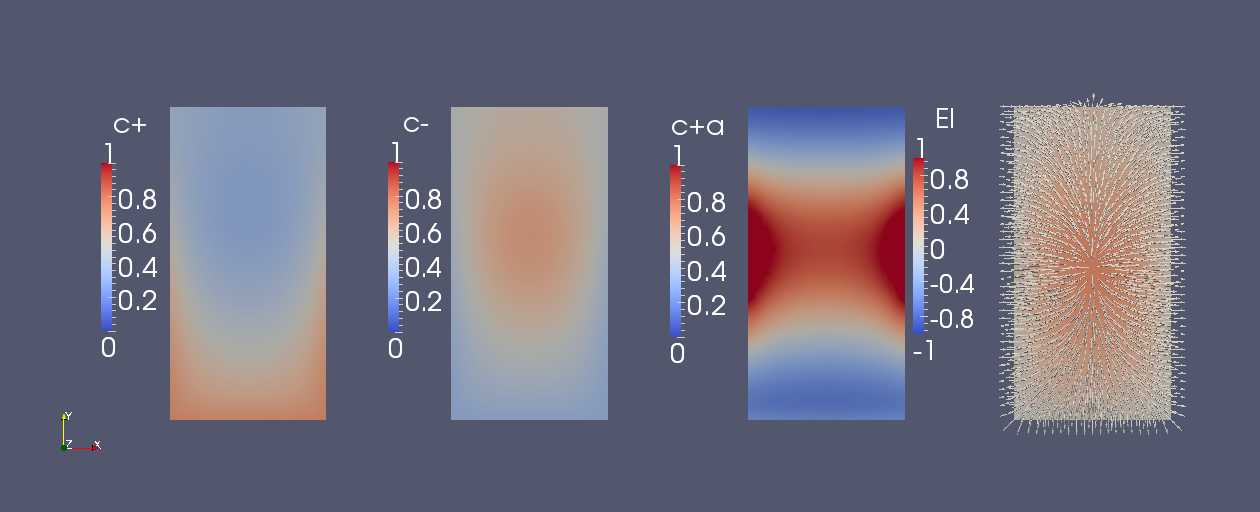} 
        \end{center}
        \vspace{-7.0mm}
        \caption{Plot of the solutions at time $t=25.6$}\label{fig:coagES-sim0256}
       \end{figure}
 \item Finally, the dynamics come to standstill with the aggregates accumulated at the bottom, cf. \cref{fig:coagES-sim0600}. The solutions are again in the ordering of \cref{fig:coagES-sim0008}.
       \begin{figure}[H]
        \begin{center}
         \includegraphics*[bb=50 40 1200 420, scale=0.32]{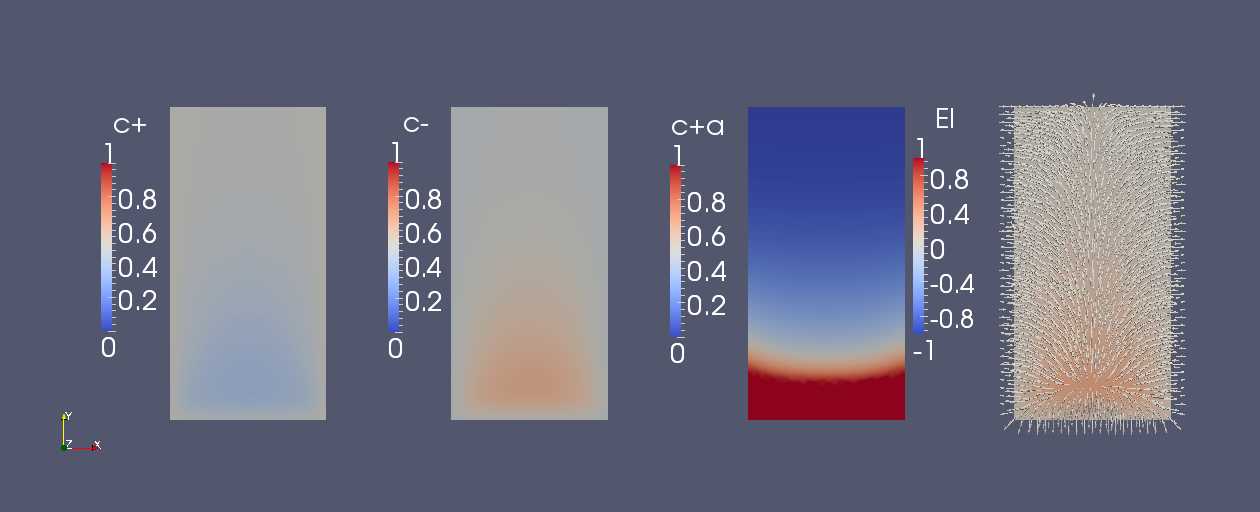} 
        \end{center}
        \vspace{-7.0mm}
        \caption{Plot of the solutions at time $t=60$}\label{fig:coagES-sim0600}
       \end{figure}
\end{enumerate}
This numerical simulation was carried out in the software HyPHM, cf. \cite{HyPHM}, which is a finite element library based on MATLAB, cf. \cite{MATLAB}.
More precisely, we applied Gummel-iteration, cf. \cite{Gummel1964}, to linearize the nonlinear system~\eqref{eq:coagES-model-poisson}--\eqref{eq:coagES-model-momBalanceTot}.
The resulting sequence of linearized systems, we discretized by Raviart-Thomas element of lowest order in space and by the implicit Euler scheme in time, cf. \cite{brezzi-book, Quarteroni-book, RaviartThomas}.
For further details to numerical investigations of \pnp{s}, we refer, e.g., to \cite{ChenCockburn_pnp, wohlmuth_pnp, GajewskiEtAl_pnp_sim, Furhmann-pnp, FrankRay2011_pnp, RayFrankNoorden_pnp}.
%
%
\section{Conclusion}
In this paper we presented a model for coagulation in electrolyte solutions. In \cref{sec:coagES-start}, we started with the well-known \nspnp\
as model for electrolyte solutions. We chose this model for electrolyte solutions, since it is a thermodynamical consistent continuum model, which captures the coupled interplay of
long-range electrostatic interaction, hydrodynamics, and mass transport. Moreover, we used this continuum model, as continuum models naturally account for many-body effects,
which are important for a better understanding of coagulation. In this paper, we proposed to include coagulation phenomena in the \nspnp\ by modeling coagulation in terms of a reaction rate~$\Rl[]^{coag}$.
For this coagulation rate~$\Rl[]^{coag}$, we derived formula~\eqref{eq:coagES-model-reactionRate} in \cref{sec:coagES-kinetics}. This formula included both, the energetic picture of coagulation according
to DLVO-theory, and the kinetic picture of coagulation in terms of the stoichiometry~\eqref{eq:coagES-kinetics-ansatzCoagKinetics}.
Secondly, by means of \eqref{eq:coagES-model-reactionRate}, we connected the atomistic DLVO-theory with the continuum \nspnp. Thereby, we combined two classical theories for electrolyte solutions,
which actually have been developed on different spatial scales. In \cref{sec:coagES-model}, this finally resulted in the model~\eqref{eq:coagES-model-poisson}--\eqref{eq:coagES-model-momBalanceTot},
which contained the microscopic picture condensed in the critical coagulation concentration ($c.c.c.$). Moreover, we demonstrated that the presented model for coagulation even captures most
of the relevant systems, if we confine ourselves to situations with vanishing barycentric flow. Finally, in this case we presented numerical computations, which showed that the resulting model
is capable to produce the expected dynamics of coagulation in electrolyte solutions. In conclusion, this paper presented a first step towards the modeling of coagulation in electrolyte solutions
based on micro-macro continuum models.
\medskip
\par
Finally, we note that there are two generalizations of the  presented model and the presented computations. Firstly, to clearly distinguish
between the dissolved particles and the aggregates, we could reformulate the above model as multiphase model in the sense that we treat the
aggregates as solid phase, whereas the fluid phase consists of the solvent and the solutes. Such a two-phase approach renders the illustrative but limited
interpretation of $\cpos$ and $\capos$ as size classes superfluous. Secondly, the numerical computations can be extend to multiscale numerics in the sense that the
critical coagulation concentration ($c.c.c.$) can be locally computed at each degree of freedom by solving suitable microscopic problems. This would fully
reveal the micro-macro character of the presented model. Both of these generalizations are currently under preparation and will be presented in forthcoming articles.
%
%
\section*{Acknowledgements}
M. Herz was supported by the Elite Network of Bavaria.
%
%
\addcontentsline{toc}{part}{References}
\fancyhead[L]{References}
\fancyhead[R]{}
\printbibliography
\end{document}